\documentstyle[12pt]{article}
\begin{document}
\begin{center}
   {\bf  SPONTANEOUS DECAY OF THE EFFECTIVE COSMOLOGICAL CONSTANT}\\
\mbox{}

                    Murat \"Ozer  and  M. O. Taha\\
               Department of Physics, College of Science\\
                  King Saud University, P.O. Box 2455\\
                         Riyadh 11451, Saudi Arabia\\
\mbox{}
                     
                             {\bf ABSTRACT}\\
\end{center}

We discuss the notion that quantum fields may induce an
effective  time  -  dependent cosmological  constant   which
decays  from  a large initial value. It is shown  that  such
cosmological  models  are  viable  in  a  non  -  de  Sitter
spacetime.
\vspace{2cm}

1. {\it Introduction.}  The cosmological  constant $\lambda$    in  the
Einstein field equations
\begin{equation} 
R_{\mu\nu}-\frac{1}{2}g_{\mu\nu}R=-8\pi GT_{\mu\nu}+\lambda g_{\mu\nu}       \end{equation}                       
corresponds  to  the energy density for the vacuum.  In  the
standard   homogeneous and isotropic model of  the  universe
described by the \mbox{Robertson - Walker} (RW) metric
\begin{equation} 
 ds^{2}=-dt^{2}+a(t)^{2}\left[\frac{dr^{2}}{1-kr^{2}}+r^{2}sin^{2}\theta d\phi^{2}\right] ,                                    
\end{equation} 
where  $k$ is the curvature constant and $a(t)$ is the scale factor
of the universe, the field equations (1) give
\begin{equation} 
\left(\frac{\dot{a}}{a}\right)^{2}=\frac{8\pi G}{3}\rho-\frac{k}{a^{2}}+\frac{\lambda}{3} ,
\end{equation} 
where $\rho$ is the density of the universe. The controversy  over
the  present value of the Hubble constant \footnote{ We denote the {\it present} value of a quantity by the subscript $p$.}$H_{p}=\left(\frac{\dot{a}}{a}\right)_{p}$ continues with the
two  recently announced values of [1] $H_{p}=80\pm17kms^{-1}Mpc^{-1}$  and [2] $H_{p}=52\pm8kms^{-1}Mpc^{-1}$. The  present
value of  the matter density $\rho_{Mp}$ lies between $10^{-48}GeV$  and $10^{-46}GeV$  while the critical density $\rho_{p}^{c}=3H_{p}^{2}/8\pi G\sim 10^{-47}GeV$. Eq.(3) can be written as
\begin{equation} 
\frac{k}{H_{p}^{2}a_{p}^{2}}=\Omega_{p}+\Omega_{p}^{V}-1 ,                   \end{equation}         
where $\Omega_{p}$ and $\Omega_{p}^{V}$ are, respectively,  the present values of the density parameter $\Omega=\rho/\rho^{c}$   and the  vacuum density  parameter $\Omega^{V}=\rho^{V}/\rho^{c}=\lambda/3H^{2}$  where $\rho^{V}=\lambda/8\pi G$. Since $H_{p}^{2}a_{p}^{2}$  is of order unity it follows that  the \mbox{right - hand} side of (4) is either $0$ (for $k=0$) or of order unity (for $k=\pm1$). The present energy density estimates give
$\Omega_{p}\sim 0.1-0.4$  from  which  we  conclude that $\mid \Omega_{p}^{V}\mid \leq 1$.  This  implies $\mid \rho_{p}^{V}\mid \leq 10^{-47}GeV$.
However,  gauge field theories imply that the value  of  the
vacuum  energy  density  $\rho^{V}$ was quite different  in  the  early
universe from its present value. For example, when the gauge
symmetry $SU(2)_{L}\times U(1)_{Y}$ of the standard  electroweak  theory
is  broken  by  the  vacuum expectation  value  of  a  Higgs
doublet,   the   energy  density  stored   in   the   broken
vacuum is  $\rho^{V}=-\frac{1}{8}M_{H}^{2}\langle\phi\rangle^{2}\sim
-3\times 10^{8}GeV^{4}$ for a Higgs boson with mass $M_{H}\approx 200GeV$ and $\langle \phi \rangle= 246GeV$. This  is  $55$   orders of magnitude larger than  the  present bound.  There  may have occurred other spontaneous  symmetry breakings  in  the  early universe; the most  celebrated  of
which  is the grand unification symmetry breaking at a  mass
scale of $M\sim 10^{16}GeV$. The vacuum energy density associated
with  this  breaking is $\sim 10^{64}GeV$ which  is  $111$  orders  of
magnitude larger than the present vacuum energy density. The
cosmological constant problem [3,4] is then to explain these
huge  orders  of magnitude differences between $\rho_{early}^{V}$  and $\rho_{p}^{V}$  in  a natural  way; in particular  without fine tuning the  values of parameters to very many decimal places.

There have been various attempts at solving the problem
of  the cosmological constant  some of which are reviewed in
Ref.[4]. One line of attack that has not been reviewed  in
Ref.[4], is suggested in several studies [5 - 7] of quantum
field  theory in de Sitter space. These studies  have  found
that  the  vacuum expectation value of the \mbox{energy - momentum}
tensor $\langle T_{\mu\nu}\rangle$  of quantum fields in de Sitter space
takes the form of  an effective cosmological energy density times the metric 
tensor. Due to quantum instabilities  the vacuum energy density  (from  now  on  we will  denote $\rho^{V}$  by $\Lambda$) decays spontaneously and  can  dominate over  the matter content of the universe. This idea has been
examined  in  Ref.[8]  which  concludes  that  such  quantum
instabilities   are   unlikely  to  provide   an   effective
cosmological energy density that decays from a large initial
value.  The  analysis  is based on the assumption  that  the
spacetime   metric  is approximately de Sitter.  This  is  a
necessary condition for the decay mechanism of Refs.[5 -  7]
to work. It is argued in Ref.[8] that this assumption cannot
be  avoided  in a model that possesses a decaying  effective
vacuum   energy   density   -  or   equivalently   effective
cosmological constant.

Shortly  after the publication of Ref.[8]  however,  a
decaying effective cosmological constant model was presented
in Refs.[9]. Since then many extensions and generalizations
of   this model  have been proposed [10 - 15]. Despite  all
the  interest  in  this  model  and  its  extensions, an
explanation  of  how any decaying - $\Lambda$ model circumvents  the
rather strong conclusions of Ref.[8] has not appeared in the
literature. It  is our purpose in this paper (1)  to  exhibit
how the negative result of  Ref.[8] has been avoided  in the
model  of  Ref.[9] and (2) to argue that decaying - $\Lambda$ models
may arise in field theoretic contexts.

Before introducing our analysis we would like to point out that, in the context of this paper, the term {\it "effective  cosmological  constant"}
means the value of $T_{00}$ for a scalar field $\phi$ that represents the vacuum in a RW  metric. This is the time - dependent vacuum energy density in this model. In a general quantum field theory in curved space - time one should calculate the effective potential and minimize it with respect to the matter fields to obtain the vacuum state. The value of the effective potential in this vacuum state is the effective cosmological constant. It should then be possible to directly relate this to the covariant vacuum energy momentum tensor, since both are defined in terms of the effective action in the vacuum state.

2. {\it Formulation  of  the  Problem.}  According  to  the
findings of Refs.[5 - 7], the quantum instabilities  in $\langle T_{\mu\nu}\rangle$  of the  quantum fields which arise in a de Sitter space provide a  variable  cosmological constant which effectively  solves
the problem. To ensure a de \mbox{Sitter - like} universe which  is
a  necessary  condition for the \mbox{gravitationally  -  mediated}
decay  mechanism to work, Ref.[8] imposes the following  two
conditions  on $\Lambda$:  (1) $\Lambda$  decays  slowly  and  (2) $\Lambda$ always dominates  over  the  matter and radiation  content  of  the
universe. It is then shown that, it seems improbable because
of  these two requirements that a realistic cosmology can be
constructed  which allows $\Lambda$ to decay from an initially  large
value.

To demonstrate the existence of decaying - $\Lambda$  cosmologies
which  evade the conditions mentioned above, we will examine
them   in  the  simplest  \mbox{variable  - $\Lambda$   model}  presented  in
Refs.[9].  This  model displays a decaying  effective  vacuum
energy  density  (or  equivalently  a  decaying  effective
cosmological  constant) in a \mbox{non - de Sitter}  context.  The
universe is neither born into a de Sitter phase nor does  it
evolve  towards  one. The model is based on  the  assumption
that,  in a general RW metric, the \mbox{energy - momentum}  tensor
$T_{\mu\nu}^{(U)}$ of the universe includes, besides $T_{\mu\nu}^{(m)}$ of matter, an additional piece $T_{\mu\nu}^{(V)}$  given by
\begin{equation} 
T_{\mu\nu}^{(V)}=-\Lambda(t)g_{\mu\nu} .
\end{equation}
The field equations then involve $\Lambda(t)$. One further specifies
the  model  by  requiring that the  energy  density  of  the
universe  is  always  given by its critical  value $\rho^{c}=3H^{2}/8\pi G$. This determines the intrinsic curvature $k$ to be unity and  yields
a definite $\Lambda(a)$:
\begin{equation} 
\Lambda=\frac{3}{8\pi G}\frac{1}{a^{2}} .
\end{equation}
In  the  radiation  era, $p=\frac{1}{3}\rho$    and it  is  possible  to  solve explicitly for $a(t)$:
\begin{equation} 
a(t)^{2}=a_{0}^{2}+t^{2} ,
\end{equation}
where $a_{0}$ is the scale factor of the universe at $t=0$. Thus  the universe is born with an initial value $\Lambda_{0}=3/8\pi Ga_{0}^{2}$   for the vacuum energy density $\Lambda(t)$ which subsequently decays so  that $\Lambda(t)\sim 3/8\pi Gt^{2}$  for $t\gg a_{0}$. The equation for $\dot{\Lambda}$ as a function of $\Lambda$ is
\begin{equation} 
\dot{\Lambda}=-2\left(\frac{8\pi G}{3}\right)^{1/2}\Lambda^{3/2}(1-\Lambda/\Lambda_{0})^{1/2} .
\end{equation}                                                    

For $t\gg a_{0}$ (i.e. $a\gg a_{0}$) $\dot{\Lambda}$   may  be  well approximated by $\dot{\Lambda}\approx -\alpha \Lambda^{3/2}$  which is  Eq.(2)  of Ref.[8]  with  $\alpha=2\sqrt{8\pi G/3}$.

It  is clear from Eq.(7) that the spacetime metric  is
not de Sitter. The particular form of $\Lambda(a)$ in Eq.(6), which  is
charecteristic of this model and its generalizations, avoids
these constraints. The resulting cosmology is realistic  and
phenomenologically viable [9]. The two conditions (1) $\Lambda \gg \rho$    for all $t$ and  (2) $\mid \dot{\Lambda}a/\Lambda\dot{a}\mid \ll 1$, imposed in Ref.[8] as de Sitter conditions,  are  therefore not satisfied. In fact   $\Lambda \gg \rho$ holds  only  when $a \approx a_{0}$ while $\mid \dot{\Lambda}a/\Lambda\dot{a}\mid=2$. One may thus conclude that the negative result  of Ref.[8]  is  indeed limited to de \mbox{Sitter - like} metrics  and does not extend to a general RW metric.

It thus remains to consider whether models of this type
may arise in a field theoretic context.  Consider a model in
which  the  evolution  of the universe  is  described  by  a
uniform scalar field $\phi$. The metric is  RW with $k=1$ and the
Lagrangian density is specified as
\begin{equation} 
{\cal L}=-\frac{R}{16\pi G}-\frac{1}{2}\partial_{\mu}\phi\partial^{\mu}\phi-V(\phi) ,
\end{equation}
where
\begin{equation} 
V(\phi)=\left\{
\begin{array}{ll}
U(\phi)              &, \qquad 0 \leq \phi \leq \phi_{1}\\
V_{1}e^{-\gamma\phi} &, \qquad \phi \geq \phi_{1}
\end{array}
\right.
\end{equation}
with $\gamma=-U'(\phi_{1})/U(\phi_{1})>0$ and $V_{1}=U(\phi_{1})e^{\gamma\phi_{1}}$ [16]. The potential $U(\phi)$  may be chosen to have  the characteristics required to spontaneously break the symmetry in $\phi \leq \phi_{1}$.  We  are interested in the large  $t$  behavior  of  the model. For $\phi \geq \phi_{1}$, the field equations
\begin{equation} 
\left(\frac{\dot{a}}{a}\right)^{2}=\frac{8\pi G}{3}(\frac{1}{2}\dot{\phi}^{2}+V(\phi))-\frac{1}{a^{2}} ,
\end{equation}
and
\begin{equation} 
\ddot{\phi}+3\frac{\dot{a}}{a}+V'(\phi)=0 ,                                  \end{equation}                 
admit the exact \mbox{non - de Sitter} solution
\begin{equation} 
a(t)=a_{2}(1+Ct)\hspace{3mm};\hspace{3mm}\phi(t)=\phi_{2}+2\gamma^{-1}ln(1+Ct) ,
\end{equation} 
where
\begin{equation} 
C^{2}=\frac{3}{8\pi G}\gamma^{2}a_{2}^{-2}(6-\frac{3}{8\pi G}\gamma^{2})^{-1}\hspace{3mm};\hspace{3mm}\phi_{2}=\gamma^{-1}ln\left[\frac{2\pi G}{3}V_{1}a_{2}^{2}(6-\frac{3}{8\pi G}\gamma^{2})\right] ,
\end{equation}            
provided  that $\gamma^{2}/8\pi G<2$.  The constant rate $\dot{a}$ of the  expansion  is less  than  unity when $\gamma^{2}/8\pi G<1$.  The solution  (13)  is  valid  for $t \geq t_{1}$ where $\phi(t_{1})=\phi_{1}$. Eqs.(14) determine $C$ and $a(t_{1})$ or equivalently $C$ and $a(0)$ if $U(\phi)$
is known.

The  scalar  field $\phi$   may  be  taken  either  as  the
nonvanishing vacuum expectation value of a quantum field  or
as  a  representation of the \mbox{non - trivial} vacuum of  curved
spacetime.  The induced effective vacuum energy  density  in
either case is $\frac{1}{2}\dot{\phi}^{2}+V(\phi)$. This yields
\begin{equation} 
\Lambda(t)=6C^{2}\gamma^{-2}(1+Ct)^{-2} \sim 6\gamma^{-2}t^{-2} ,
\end{equation}   
for  large  $t$. Thus $\Lambda(t)$ decays to zero from any initially large
value it might have had. One observes that $\dot{\Lambda}=-\sqrt{2/3}\,\gamma\Lambda^{3/2}$  as in Eq.(2) of
Ref.[8]  while  $\mid \dot{\Lambda}a/\Lambda\dot{a}\mid=2$  as above. For large $t$  the scale  factor $a(t)$ is  not  much  different from that of the  critical  density model   indicating  that  the  two  models  have   desirable cosmological features in common.

3. {\it Stability Analysis.} Before we discuss the stability
of  our  solution, an important and novel feature of  models
with $\rho_{\phi}$ (or $\Lambda$)  decaying  as $a^{-2}$ should  be pointed out. Using  Eqs.(13)and (14), Eq.(11) can be cast into
\begin{equation} 
\left(\frac{\dot{a}}{a}\right)^{2}=-\frac{k_{eff}}{a^{2}} ,
\end{equation}                                              
where
\begin{equation} 
k_{eff}=k-\frac{1}{\left(1-\frac{\gamma^{2}}{2\chi^{2}}\right)}<0 ,
\end{equation}                                         
with $k=1$ and $\chi^{2}=8\pi G$. The parameter $k_{eff}$ is the effective curvature scalar  of the  present model universe. Thus, even though the  \mbox{three  - geometry}  of  the universe is closed $(i.e.\, k=1)$,  the
universe evolves as if it is open due to the fact that  $k_{eff}<0$.
This   always  happens  even  with  more  realistic  models
containing  radiation  and matter  whenever $\rho_{\phi}$ (or $\Lambda$)  decays as $a^{-2}$. Consequently  the  universe  avoids  collapse  as  long  as $k_{eff}$ remains   negative. Hence the intriguing  possibility  of  a
closed but \mbox{ever - expanding} universe.

Stability of the system of equations (11) and (12) with $V(\phi)$
given  in Eq.(10) can be investigated by introducing  a  new
time  variable $\tau$  and defining a new scale factor $\alpha(\tau)$  such  that [17] (see also Ref.[18])
\begin{equation} 
\begin{array}{l}
\frac{d}{dt}=V^{1/2}\frac{d}{d\tau} \vspace{3mm}\\
a(\tau)=exp(\alpha(\tau)) .
\end{array}
\end{equation}
In  terms of these variables equations  (11) and (12) become
respectively
\begin{equation} 
\alpha'^{2}=\frac{\chi^{2}}{3}\left(\frac{1}{2}\phi'^{2}+1\right)-\frac{ke^{-2\alpha+\gamma\phi}}{V_{1}}
\end{equation}
\begin{equation} 
\phi''=\frac{1}{2}\gamma\phi'^{2}-3\alpha'\phi'+\gamma
\end{equation}
where a prime denotes differentiation with respect to $\tau$.  An
equation  which  does  not  depend   on $k$ and $V_{1}$ is  obtained   by
differentiating Eq.(19):
\begin{equation} 
\alpha''=\frac{1}{2}\gamma\alpha'\phi'-\frac{\chi^{2}}{3}(\phi'^{2}-1)-\alpha'^{2} .
\end{equation}
Defining the new variables $x=\phi'$ and $y=\alpha'$  the  system of equations   (20) and (21) are   transformed  into  the  following  \mbox{two  -  dimensional} autonomous system
\begin{equation} 
\begin{array}{l}
x'=\frac{1}{2}\gamma x^{2}-3xy+\gamma ,\vspace{3mm}\\
y'=\frac{1}{2}\gamma xy-\frac{\chi^{2}}{3}(x^{2}-1)-y^{2} .
\end{array}
\end{equation}
The system (22) possesses the following critical points
\begin{eqnarray} 
(x,y)& = & \left(\pm 1 , \pm \frac{\gamma}{2}\right) ,\\
(x,y)& = & \left(\pm \frac{\gamma\sqrt{2}}{\sqrt{6\chi^{2}-\gamma^{2}}} , \pm \frac{\chi^{2}\sqrt{2}}{\sqrt{6\chi^{2}-\gamma^{2}}}\right) .  
\end{eqnarray}
In  Figure 1, we depict the phase flow in the $xy$  plane   for $\chi^{2}=3$, and $\gamma=1$, where the critical points corresponding to (23) and (24) are
the  attractor  $A$ and the saddle point $S$, respectively.  The $k=+1$
solutions   are  separated  from  the $k=-1$  solutions   by   the $k=0$
trajectories   which are the  hyperbolae $y^{2}=\frac{1}{2}x^{2}+1$, as  follows  from Eq.(19).  (For a detailed discussion of the solutions  for
other  values  of   the coupling   we refer  the  reader  to
Ref.[18])  Figure  1  indicates  that  only  some  of  the $k=+1$
solutions  are  asymptotically stable.  Does   our  solution
given  in  Eq.(13)  belong  to  the  family  of  the  stable
solutions? The answer is affirmative. As we have noted above
cosmological  solutions  with $\rho_{\phi}$  decaying  as  $a(t)^{-2}$
lead  to   an effective curvature scalar given in Eq.(17). It is shown  in
Ref.[18] that irrespective of the value of the coupling  all
the $k=-1$   solutions  are  asymptotically  stable.   Hence   the
stability  of  our  solution follows  as  long  as $k_{eff}$  remains
negative, which is the case for $\gamma^{2}/\chi^{2}<2$  \footnote{The  stability of  all the cosmological solutions  with $\rho_{\phi}=\frac{1}{2}\dot{\phi}^{2}+V(\phi)$  decaying  as  $a(t)^{-2}$  and $V(\phi)=V_{1}e^{-\gamma\phi}$   can be investigated by substituting  in
Eq.(11) $\rho_{\phi}=3c/\chi^{2}a^{2}$, where $c>1$ so that $k_{eff}$ in Eq.(17) is negative. One  then finds  that all the $k=+1$  solutions with this property are indeed stable. We hope to expand on this in future  work.}

4. {\it Primordial Nucleosynthesis.} It is argued  in Ref.[8]
that  cosmological models that are dominated by  a  decaying
cosmological  constant predict a primordial $^{4}He$ abundance  today
which  is $<10^{-26}$ (Assuming that $^{4}He$  are not direct products of $\Lambda$). This  conclusion  follows from combinations  of  assumptions
which do not hold in models of the type considered here. The
details  of  primordial nucleosynthesis in  decaying  vacuum
cosmologies  with $\Lambda$ decaying as $a^{-2}$   have  been  considered  in Ref.[14].  Here ,we will outline how the method of  Ref.[14]
can  be  implemented in a \mbox{field - theoretic}  model.  As  the
scalar field $\phi$  evolves and decays into radiation and matter,
there  comes an epoch at which the scale factor is no longer
given by that in Eq.(13). This is due to the accumulation of
radiation   (i.e.   relativistic   particles)   and   matter
(nonrelativistic particles). The period starting  with  this
epoch,  at which $t=t_{\ast}$ and $a=a_{\ast}$, may be called the \mbox{radiation - dominated} era, and lasts, as in the standard model, until the radiation and matter energy densities become equal  at $a=a_{eq}$. The decay of the $\phi$ - field energy density into radiation and matter can  be
modeled by [19]
\begin {equation} 
\rho_{r}=\Gamma_{r}\dot{\phi}^{2}\hspace{3mm},\hspace{3mm}\rho_{m}=\Gamma_{r}\dot{\phi}^{2} ,                                            
\end{equation}
where $\Gamma_{r}$ and $\Gamma_{m}$   are  very slowly varying functions  of  time.  We approximate  them by constants here. Upon the  inclusion  of
radiation and matter Eq.(11) changes to
\begin{eqnarray} 
\left( \frac{\dot{a}}{a}\right)^{2} & = &\frac{8\pi G}{3}(\rho_{r}+\rho_{m}+\rho_{\phi})-\frac{1}{a^{2}}\nonumber \\
                                        & = &\frac{8\pi G}{3}\left[\left(\Gamma_{r}+\Gamma_{m}+\frac{1}{2}\right)\dot{\phi}^{2}+
V(\phi)\right]-\frac{1}{a^{2}} ,
\end{eqnarray}
and Eq.(12), which follows from the conservation equation
\begin {equation} 
d\left[(\rho_{\phi}+\rho_{r}+\rho_{m})a^{3}\right]+(p_{\phi}+p_{r}+p_{m})da^{3}=0
\end{equation}                                  
when $\rho_{r}=\rho_{m}=0$, is replaced by
\begin {equation} 
(1+2\Gamma_{r}+2\Gamma_{m})\ddot{\phi}+3\left(\frac{4}{3}\Gamma_{r}+\Gamma_{m}+1\right)\frac{\dot{a}}{a}\dot{\phi}+\frac{dV(\phi)}{d\phi}=0 . 
\end{equation}                     
Due to the creation of radiation and matter the scalar field
and its potential change to
\begin {equation} 
\phi(t)=\phi_{2}-\frac{2}{n\gamma}ln(1+Ct)\hspace{3mm},
\hspace{3mm}V(\phi)=DV_{1}e^{\gamma\phi} , 
\end{equation} 
where  the  coefficients $n$ and $D$  are to be determined from Eqs.(26)
and (28) self - consistently. We obtain the relations
\begin {equation} 
D=\frac{2}{n(n+1)}\hspace{3mm},\hspace{3mm}\Gamma_{r}+\Gamma_{m}=n\left(\frac{n-1}{n+1}\right) .
\end{equation} 
Since  the kinetic energy of the scalar field cannot convert
to radiation and matter with hundred percent efficiency, the
sum $\Gamma_{r}+\Gamma_{m}$   must be less than unity. Hence, 
it follows from Eq.(30) that $n<1+\sqrt{2}$. Without  loss of
generality, we will  take   in  the following $n=2$. Next, we express the scalar field energy  density $\rho_{\phi}$ and  the  pressure 
$p_{\phi}$  in terms of $\Gamma_{r}+\Gamma_{m}$   and make the  assumption
that  even  though  radiation  and  matter  are  in  thermal
equilibrium they evolve almost independently of each  other,
as  in the standard model. Upon multiplying through by $a$  the
energy   conservation  equation  (27)  then  leads  to   two
approximately independent equations:
\begin {equation} 
d(\rho_{r}a^{4})=\frac{21\Gamma_{r}}{\chi^2(6-3\gamma^{2}/\chi^{2})}ada , 
\end{equation} 
\begin {equation} 
d(\rho_{m}a^{3})=\frac{21\Gamma_{m}}{\chi^2(6-3\gamma^{2}/\chi^{2})}da . 
\end{equation} 
We  assume that Eqs.(31) and (32) hold approximately  during
the   entire  evolution  of  the  universe,  probably   with
different  values  for $\Gamma_{r}$ and $\Gamma_{m}$ in different periods.  Integrating these equations from $a$ to $a_{eq}$  results in
\begin {equation} 
\rho_{r}^{RD}=\frac{\omega_{eq}^{(r)}a_{eq}^{2}\chi^{-2}}{a^{4}}+
\frac{\frac{21}{2}\Gamma_{r}\chi^{-2}(6-3\gamma^{2}/\chi^{2})^{-1}}{a^{2}} ,
\end{equation} 
\begin {equation} 
\rho_{m}^{RD}=\frac{\omega_{eq}^{(m)}a_{eq}^{2}\chi^{-2}}{a^{3}}+
\frac{21\Gamma_{m}\chi^{-2}(6-3\gamma^{2}/\chi^{2})^{-1}}{a^{2}} ,
\end{equation} 
where $\omega_{eq}^{(r)}=\chi^{2}\rho_{eq}a_{eq}^{2}-\frac{21}{2}\Gamma_{r}/(6-3\gamma^2/\chi^{2})$ and  $\omega_{eq}^{(m)}=\chi^{2}\rho_{eq}a_{eq}^{2}-21\Gamma_{m}/(6-3\gamma^2/\chi^{2})$.  Due  to  the fact that majority of  the  existing
particles  are  relativistic and that the  scale  factor  is
relatively  small  in the \mbox{radiation - dominated}  era,  the
$a^{-4}$ term  in Eqs.(33) dominate over the $a^-{2}$ term. Thus, to a  very
good  approximation  near  and  at $T \approx 1MeV$   the  radiation  energy
density reduces to
\begin {equation} 
\rho_{r}^{RD} \approx \frac{\omega_{eq}^{(r)}a_{eq}^{2}\chi^-{2}}{a^{4}} .
\end{equation} 
It  should  be  noted that the radiation energy  density  in
Eqs.(35) has the same dependence on the scale factor  as  in
the  standard model. Substituting Eq.(35) into  Eq.(25)  and
neglecting $\rho_{m}$ and $\rho_{\phi}$, the scale factor is found to be
\begin {equation} 
a(t) \approx \left(\sqrt{\frac{4}{3}\omega_{eq}^{(r)}}a_{eq}(t-t_{\ast})+
a_{\ast}^{2}\right)^{1/2} ,
\end{equation}                      
where $t_{\ast}$ at which $a=a_{\ast}$  marks  the  epoch when the scalar  field  and  the radiation  energy  densities  are  equal.  The  universe  is
assumed to enter the radiation - dominated era once $a>a_{\ast}$,  and
$a \gg a_{\ast}$ near  and  at $T \approx 1MeV$. Hence, the behavior of the scale factor  is $a(t) \sim t^{1/2}$, just as in the standard model. The rest of the analysis  of primordial  nucleosynthesis  is  similar  to  that  in   the standard  model. The details can be found in Ref.[14]  where
it  is  shown  that the desired  $^{4}He$ abundance is  obtained  in
models with a vacuum energy decaying as $a^{-2}$.

5. {\it Conclusions.} The existence of decaying - $\Lambda$ models,
in  phenomenological or field theoretic contexts,  indicates
that the negative result of Ref.[8] hinges on the assumption
that  spacetime is de \mbox{Sitter - like}. Our conclusion is  that
it  is not unlikely that a realistic quantum field cosmology
with  a  decaying  effective cosmological  constant  can  be
constructed.  There is no general argument  against  such  a
construction. It is only necessary that the  spacetime  be
\mbox{non  - de Sitter} in such a theory. Cosmological models based
on  a homogeneous  scalar field whose energy density behaves
like  a time - variable cosmological constant in a \mbox{non -  de
Sitter} spacetime have long been considered. References  [16]
, [20 - 22], and [23]  are examples of such an endeavor.

{\bf References}
                        
1. W. Freedman et al., Nature {\bf 371},757 (1994).\\
2. A. Sandage et al. ApJ, {\bf 423}, 16 (1994).\\
3. L. Abbott, Sci. Am., {\bf 258}, 106 (1988).\\
4. S. Weinberg, Rev. Mod. Phys., {\bf 61}, 1 (1989).\\
5. N. Myhrvold, Phys. Rev. {\bf D28}, 2439 (1983).\\
6. L. Ford, Phys. Rev., {\bf D31}, 710 (1985).\\
7. E. Mottola, Phys. Rev., {\bf D31}, 754 (1985).\\
8. W. A. Hiscock, Phys. Lett., {\bf 166B}, 285 (1986).\\
9. M. \"Ozer and M. O. Taha, Phys. Lett., {\bf 171B}, 363 (1986)
; Nucl. Phys., {\bf B287}, 776 (1987).\\
10.  M. Gasperini, Phys. Lett., {\bf B194}, 3406 (1987).\\
11. W. Chen and Y. S. Wu, Phys. Rev. {\bf D41}, 695 (1990) ;
        Erratum Phys. Rev. {\bf D45}, 4728 (1992).\\
12. D. Pavon Phys. Rev. {\bf D43}, 375 (1991).\\
13. M. S. Berman, Phys. Rev. {\bf D43}, 1075 (1991).\\
14. A. M. Abdel - Rahman, Phys. Rev. {\bf D45}, 3497 (1992).\\
15. J. Matyjasek, Phys. Rev. {\bf D45}, 4154 (1995).\\
16. Similar exponential potentials have been considered  by
several authors as inspired  by string theory. See, e.g. M. \"Ozer and  M.
O. Taha, Phys. Rev. {\bf D45}, 997 (1992).\\
17.  A.  B.  Burd and J. D. Barrow, Nucl. Phys.  {\bf B308},  929 (1988).\\
18. J. J. Halliwel, Phys. Lett., {\bf B185}, 341(1987).\\
19. Creation of matter due to the decay of the scalar field
as  modeled  in Eq.(25) is similar to the mechanism  in  the
\mbox{power  -  law}  inflationary model  of   F.  Lucchin  and  S.
Matarrese, Phys. Rev. {\bf D32}, 1316 (1985).\\
20. P. J. E. Peebles and B. Ratra, ApJ. {\bf 325}, L17 (1988).\\
21.  B.  Ratra  and  P. J. E. Peebles, Phys. Rev.  {\bf D37},  3406 (1988).\\
22.  B. Ratra and A. Quillen, Mon. Not. R. Astron. Soc. {\bf 259}, 738 (1992).\\
23.  K.  M. Mubarak and M. \"Ozer, Clas. and Quant. Grav. {\bf 15}, 75 (1998).  
\newpage

FIGURE CAPTIONS:

  Figure  1:  Phase flow in the $xy$  plane for $\chi^{2}=8\pi G=3$, and $\gamma=1$.  The  dashed curves   are   the   hyperbolae $y^{2}=\frac{1}{2}x^{2}+1$ corresponding   to    the $k=0$
trajectories. The  region  between  the   two   hyperbolae
correspond to the $k=+1$ trajectories while the regions inside the
hyperbolae correspond to the $k=-1$ trajectories. The points $A$ and
$S$ are the attractor and the saddle points, respectively.

\end{document}